\documentclass[epj,referee]{svjour}
\usepackage{graphicx}
\begin{document}
\title{Estimation of experimental data redundancy and related statistics}
\author{Igor Grabec}
\institute{Faculty of Mechanical Engineering, University of Ljubljana,\\
A\v{s}ker\v{c}eva 6, PP 394, 1001 Ljubljana, Slovenia,\\ 
Tel: +386 01 4771 605, Fax: +386 01 4253 135, E-mail: igor.grabec@fs.uni-lj.si \\}
\date{Received: date / Revised version: date}

\abstract{Redundancy of experimental data is the basic statistic from which the complexity of a natural phenomenon and the proper number of experiments needed for its exploration can be estimated. The redundancy is expressed by the entropy of information pertaining to the probability density function of experimental variables. Since the calculation of entropy is inconvenient due to integration over a range of variables, an approximate expression for redundancy is derived that includes only a sum over the set of experimental data about these variables. The approximation makes feasible an efficient estimation of the redundancy of data along with the related experimental information and information cost function. From the experimental information the complexity of the phenomenon can be simply estimated, while the proper number of experiments needed for its exploration can be determined from the minimum of the cost function. The performance of the approximate estimation of these statistics is demonstrated on two--dimensional normally distributed random data.
} 

\maketitle

\section{Introduction}
\label{intro}
The basic task of experimental physical exploration of natural phenomena is to provide quantitative data on measured variables and, from them extract physical laws \cite{gs}. Related to this task, experimenters must decide how many experiments to perform in order to provide proper experimental data. We know that it is reasonable to repeat experiments as long as they yield essentially new data, and to stop repetition when the data become redundant. In order to describe this concept objectively, we have introduced in previous articles \cite{ig1,ig2} two statistics called experimental information $I$ and redundancy $R$ of experimental data based on the entropy of information \cite{les}. Their difference $C=R-I$ can be interpreted as the information cost function of the experimental exploration. From the cost function minimum, the proper number $N_\circ$ of experiments can be determined in an objective way. The entropy of information is defined by the integral of a nonlinear function of the probability density function of experimental data, and consequently its calculation is numerically demanding. This property represents a serious obstacle, especially when treating multivariate data. Therefore, our aim is to show how this obstacle can be effectively avoided by estimating data redundancy without integration. For this purpose we first briefly repeat the route to the definition of redundancy \cite{ig1,ig2} and subsequently show how the integral in the corresponding expression can be approximated. The performance of the derived approximate method of calculation is demonstrated using two--dimensional normally distributed random data. 

\section{Redundancy of experimental data}

Let us consider a phenomenon characterized by $N$ measurements of a variable 
$x$ using an instrument with span $S_x =(-L,L)$. Properties of the instrument are specified by calibration on a unit $u$. The probability density function (PDF) of the instrument's output scattering during calibration is described by the scattering function $\psi(x,u)$. When the scattering is caused by mutually independent disturbances in the experimental system, the scattering function is Gaussian \cite{gs,les} : 
\begin{equation}
\psi(x,u)=g(x-x_i,\sigma)=\frac{1}{\sqrt{2\pi}\,\sigma}\exp \biggl[-\frac{(x-u)^2}{2\sigma^2}\biggr].
\label{gau}
\end{equation}
We apply this function in our further treatment. The mean value $u$ and standard deviation $\sigma$ can be estimated statistically by repetition of calibration. 

Let $x_i$ denote the most probable instrument output in the $i$--th experiment. Using $\psi (x,x_i)$ we describe the properties of the explored phenomenon during the $i$--th experiment. Similarly, the properties in a series of $N$ repeated experiments, which yield the basic data set $\{x_i ;\,i=1,\ldots,N\}$, are described by the experimentally estimated PDF:
\begin{equation}\label{mm}
f_N(x)\,=\,\frac{1}{N}\,\sum_{i=1}^N \psi (x,x_i) .
\label{pdf}
\end{equation}
In addition, we introduce a uniform reference PDF $\rho(x) = 1/(2L)$ indicating that all outcomes of the experiment are hypothetically equally probable before executing the experiments. 

Based upon functions $f_N(x)$ and $\rho(x)$ we describe the indeterminacy of variable $x$ by the negative value of the relative entropy \cite{ct,kol,ka}:
\begin{equation}
H_x=-\int_{S_x} f(x) \log \Bigl(\frac{f_N(x)}{\rho(x)}\Bigr) \,dx .
\label{Hx}
\end{equation}
Similarly, we describe the uncertainty $H_u$ of calibration performed on a unit $u$ by:
\begin{equation}
H_u=-\int_{S_x} \psi(x,u) \log \Bigl(\frac{\psi(x,u)}{\rho(x)}\Bigr) \,dx .
\label{Hu}
\end{equation}
Using the difference of these statistics we define the experimental information:
\begin{eqnarray}
I&=&H_x-H_u \nonumber\\
&=&-\int_{S_x} f(x) \log (f_N(x)) \,dx \nonumber\\
&&+ \int_{S_x} \psi(x,u) \log (\psi(x,u)) \,dx .
\label{Ix}
\end{eqnarray}
Using Eq.\,\ref{pdf} in this expression we get:
\begin{eqnarray}
I=\log (N)&-&\frac{1}{N}\sum_{i=1}^N \int_{S_x} \psi(x,x_i) \log \Bigl(\sum_{j=1}^N \psi(x,x_j)\Bigr) \,dx \nonumber\\
&-&\int_{S_x} \psi(x,u) \log \Bigl(\psi(x,u)\Bigr) \,dx.
\label{Ix2}
\end{eqnarray}
If we express the logarithm in the second term as:
\begin{equation}
\log \Bigl(\sum_{j=1}^N \psi(x,x_j)\Bigr)=\log \psi(x,x_i) + 
\log\Bigl(1 + \sum_{j\#i}^N \frac{\psi(x,x_j)}{\psi(x,x_i)} \Bigr)
\end{equation}
we obtain:
\begin{eqnarray}
I&=&\log (N)+\frac{1}{N}\sum_{i=1}^N \int_{S_x} \psi(x,x_i) \log \Bigl(\psi(x,x_i)\Bigr) \,dx\nonumber\\
&-&\frac{1}{N}\sum_{i=1}^N \int_{S_x} \psi(x,x_i) \log\Bigl(1 + \sum_{j\#i}^N \frac{\psi(x,x_j)}{\psi(x,x_i)} \Bigr) \,dx \nonumber\\
&-&\int_{S_x} \psi(x,u) \log \Bigl(\psi(x,u)\Bigr) \,dx.
\label{Ix3}
\end{eqnarray}
The second and the fourth term on the right side of this equation yield 0 and we get:
\begin{equation}
I=\log (N)-\frac{1}{N}\sum_{i=1}^N \int_{S_x} \psi(x,x_i) \log\Bigl(1 + \sum_{j\#i}^N \frac{\psi(x,x_j)}{\psi(x,x_i)} \Bigr) \,dx.
\label{Ix4}
\end{equation}
With the last term we introduce the statistic called redundancy of data: 
\begin{equation}
R=\frac{1}{N}\sum_{i=1}^N \int_{S_x} \psi(x,x_i) \log\Bigl(1 + \sum_{j\#i}^N \frac{\psi(x,x_j)}{\psi(x,x_i)} \Bigr) \,dx
\label{R}
\end{equation}
with which we get the basic relation:
\begin{equation}
I=\log (N)- R
\label{IR}
\end{equation}
If $\vert x_i-x_j\vert \gg\sigma$ for all pairs $i\#j$, there is no overlapping of functions 
$\psi(x,x_i),\psi(x,x_j)$; therefore, the sum in the logarithm is $\sim 0$, and consequently the redundancy is $R\sim 0$. In the opposite case, when 
$\vert x_i-x_j\vert \ll\sigma$, it follows that $\psi(x,x_i)\sim\psi(x,x_j)$. Due to good overlapping in this case, the corresponding term in the expression of $R$ yields $\log(2)/N$ and $R>0$. 

This property indicates that experimental information is increasing with increasing $N$ as 
$I\sim\log (N)$ if the acquired data are well separated with respect to $\sigma$. However,  
with an increasing number of data, they are ever more densely distributed, which results in an increasing overlapping of distributions that causes increasing redundancy of measurements. Although the expression in Eq.\,\ref{R} for redundancy $R$ is rather cumbersome due to the included integral, we expect that $R$ could be estimated without integration by the simpler function of distances between data points. For this purpose we next consider the properties of the scattering function $\psi(x,x_i)$. 

If the Gaussian function $\psi(x,x_i)=g(x-x_i,\sigma)$ is considered as an approximation of the delta function $\delta(x-x_i)$, and the logarithm as a slowly changing function, the integration in Eq.\,\ref{R} can be carried out, which yields for the redundancy the first order approximate expression without the integral:
\begin{equation}
R_1\sim\frac{1}{N}\sum_{i=1}^N  \log\biggl\{1 + \sum_{j\#i}^N \frac{\psi(x_i,x_j)}{\psi(x_i,x_i)}. \biggr\} 
\label{Rap1}
\end{equation}
If we take into account Eq.\,\ref{gau}, we get for the redundancy the following approximate expression that depends only on standard functions of distances between data points:
\begin{equation}
R_1\sim\frac{1}{N}\sum_{i=1}^N  \log \biggl\{ 1 + \sum_{j\#i}^N \exp \Bigl[-\frac{(x_i-x_j)^2}{2\sigma^2}\Bigr]\biggr\} 
\label{Rap2}
\end{equation}

However, this first order approximation is rather rough because the distribution $\psi(x_i,x_j)$ has the width $\sigma>0$ and the logarithm in Eq.\,\ref{R} includes the fraction of functions $\psi(x,x_j)/\psi(x,x_i)$. To proceed to a better approximation, we have examined the case of just two data points, since it mainly determines the property of the redundancy. In this case the integration of the first three terms in a Taylor series expansion of the logarithm yields the second approximation:
\begin{equation}
R_2\sim\frac{1}{N}\sum_{i=1}^N  \log \biggl\{ 1 + \sum_{j\#i}^N \exp \Bigl[-\frac{(x_i-x_j)^2}{4\sigma^2}\Bigr]\biggr\},
\label{Rap3}
\end{equation}
which is obtained from the previous one by merely changing $2\sigma^2\rightarrow 4\sigma^2$. This property indicates that a still better approximation could be obtained by properly adapting 
$2\sigma^2$ in Eq.\,\ref{Rap2}. For this purpose we have proceeded with numerical investigations which have shown that a nearly optimal approximation is obtained if $2\sigma^2$ in Eq.\,\ref{Rap2} is replaced by $\sim 5.1\sigma^2$:  
\begin{equation}
R_o\sim\frac{1}{N}\sum_{i=1}^N  \log \biggl\{ 1 + \sum_{j\#i}^N \exp \Bigl[-\frac{(x_i-x_j)^2}{5.1\sigma^2}\Bigr]\biggr\}. 
\label{Rap4}
\end{equation}
Numerical investigations have further shown that this formula also yields good results in cases with many data points. 

Since the integral is excluded from Eq.\,\ref{Rap4}, the redundancy $R$ can be estimated from Eq.\,\ref{Rap4} with essentially less computational effort than from Eq.\,\ref{R}. This advantage is especially outstanding in a multivariate case where the redundancy is defined by multiple integrals, while in the approximate formula in Eq.\,\ref{Rap4} only the term $(x_i-x_j)$ in the exponential function has to be replaced by the norm of corresponding vectors. Due to this advantage, it is also reasonable to estimate approximately the experimental information using the basic formula $I=\log(N)-R$. The experimental information $I$ converges with the increasing number of data $N$ to a certain limit value from which the complexity of the phenomenon under investigation can be estimated using the formula $K\approx \exp (I_{N\rightarrow \infty})$ introduced  previously \cite{ig1,ig2}. The complexity $K$ indicates how many non--overlapping scattering distributions are needed in the estimator Eq.\,\ref{fxy} to describe the PDF of the observed phenomenon. 

The information cost function is the difference of the redundancy and experimental information: 
$C=R-I$. During minimization of this cost, the experimental information provides for a proper adaptation of the PDF estimator to the experimental data, while the redundancy prevents excessive growth of the number of data points. 
By the position of the cost function minimum we introduce the proper number $N_o$ of the data and the corresponding experiments that are needed to judiciously represent the phenomenon under exploration. By inserting the expression $I=\log (N) -R$ into $C=R-I$, we obtain for the information cost function the formula:
\begin{equation}
C=2R-\log(N). 
\label{CF}
\end{equation}
Therefore the proper number $N_o$ can also be determined from the approximately estimated redundancy $R_o$. This number roughly corresponds to the ratio between the magnitude of the characteristic region where experimental data appear and the magnitude of the characteristic region covered by the scattering function \cite{ig1,ig2}. 

\section{Numerical examples}
\begin{figure}
\centering
\includegraphics[width=3.375in]{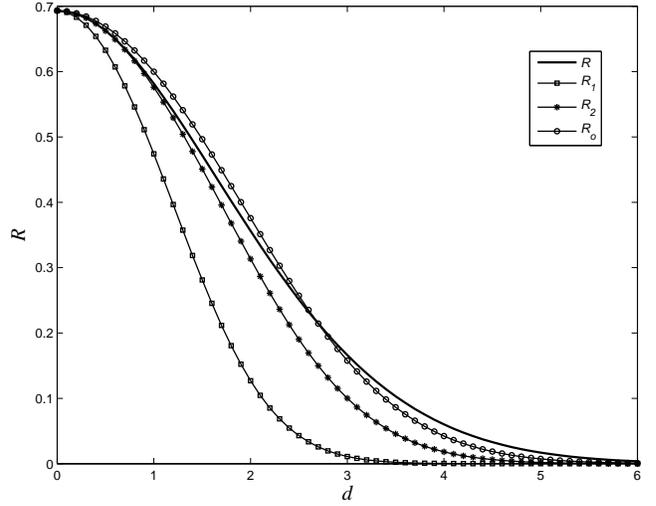}
\caption{Dependence of redundancy $R$ on relative distance $d=(x_1 -x_2)/\sigma$ between data points as determined by the integral in Eq.\,\ref{R}, and approximations in Eqs.\,\ref{Rap2},\ref{Rap3},\ref{Rap4}.}
\label{R_d}
\end{figure}

To demonstrate the properties of the approximations $R_1$, $R_2$, $R_o$ let us first consider the case of just two data points separated by a distance $x_1 -x_2$. Fig.\,\ref{R_d} shows the dependence of redundancy 
$R$ on relative distance $d=(x_1 -x_2)/\sigma$ as determined by the integral in Eq.\,\ref{R} and approximations in  Eqs.\,\ref{Rap2},\ref{Rap3},\ref{Rap4}. Improvement achieved by subsequent steps of approximation and a fairly good agreement between approximation $R_o$ and $R$ calculated by the integral is evident. However, in a case with more data points we can generally expect slightly worse agreement due to overlapping of more than two scattering functions in the sum of the approximation formula in Eq.\,\ref{Rap4}. The performance in such a case is demonstrated in the next example.   

In order to provide for reproduction of the demonstrated example, we consider a two--dimensional Gaussian random phenomenon with zero mean value. The standard deviation of both components is equal to $s=2.5$, while their covariance is zero. The data generated by a standard Gaussian generator are represented in the two-dimensional span $(-10,+10)\otimes(-10,+10)$ using the scattering width $\sigma=0.5$. In such a case we can theoretically predict that the proper number of data samples should be $N_o\approx (s/\sigma)^2=25$.
 
\begin{figure}
\centering
\includegraphics[width=3.375in]{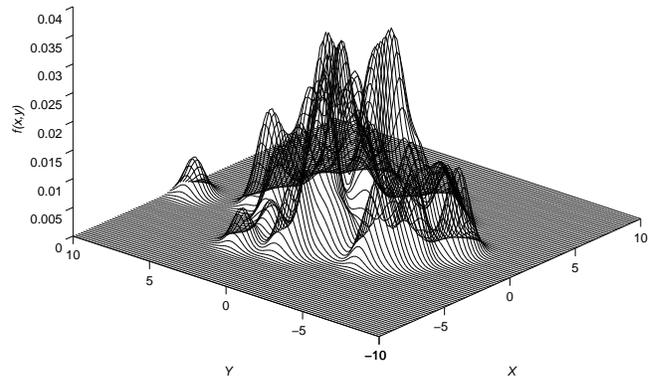}
\caption{PDF determined by 100 data points $x_i,y_i$.}
\label{fxy}
\end{figure}

For the demonstration, a set of $N_{max}=100$ two-dimensional data samples $\{(x_i,y_i); i=1\ldots N_{max}\}$ was generated. The corresponding probability density function was estimated using Eq.\,\ref{pdf} adapted to the two--dimensional case with statistically independent components:  
\begin{equation}\label{mm}
f_N(x,y)\,=\,\frac{1}{N}\,\sum_{i=1}^N \psi (x,x_i)\psi (y,y_i) .
\label{pdf2}
\end{equation}
The resulting PDF with $N=100$ is graphically represented in Fig.\,\ref{fxy}. 

\begin{figure}
\centering
\includegraphics[width=3.375in]{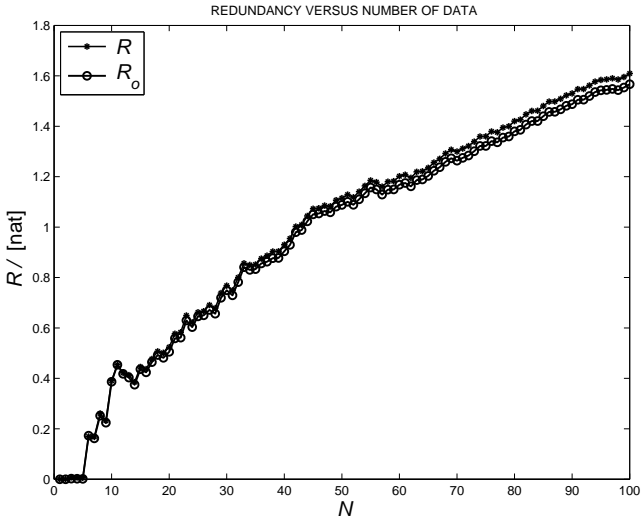}
\caption{Dependence of redundancy $R$ on number $N$ of data points as determined by the integral in Eq.\,\ref{R} -- ($R$), and approximation in Eq.\,\ref{Rap4} -- $(R_o)$ adapted to the two--dimensional case.}
\label{RaN}
\end{figure}

From the generated data the redundancy was calculated using Eqs.\,\ref{R} and \ref{Rap4} adapted to the two--dimensional case. The dependence of redundancy $R$ on the number $N$ of accounted data points is shown in Fig.\,\ref{RaN}. Fairly good agreement between both statistics is again evident. 

\begin{figure}
\centering
\includegraphics[width=3.375in]{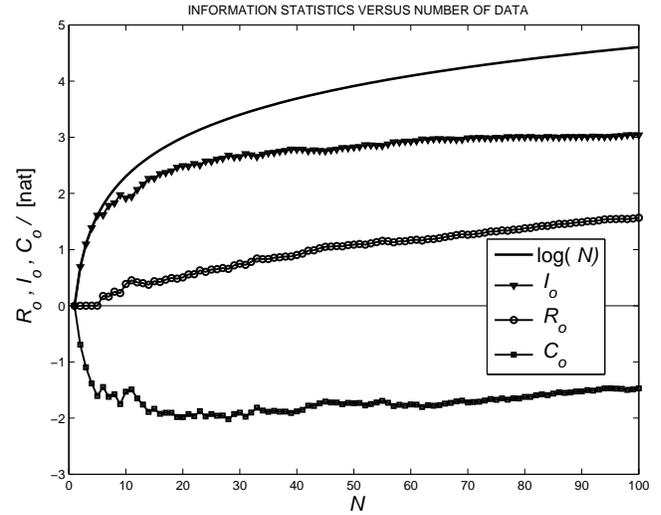}
\caption{Dependence of information statistics on the number $N$ of data points as approximately determined from Eq.\,\ref{Rap4}. The minimum of the cost function occurs at $N=28$.}
\label{IRCaN}
\end{figure}

Approximately estimated redundancy was further utilized in the calculation of statistics $I$ and $C$. They are shown as functions of the number of data points $N$ in Fig.\,\ref{IRCaN} together with $R(N)$ and $\log(N)$. Agreement with the same statistics calculated more exactly by integration can be established by comparing this figure with Fig.\,\ref{IRCaN}. In both cases we obtain for the proper number the value $N_o=28$. This value depends on the statistical properties of the data set used in its calculation; a statistical estimation from $100$ different data sets yields the estimate $N_o\approx 25 \pm 13$ which agrees well with the theoretically predicted value $N_o=25$. Similarly as in the one--dimensional case \cite{ig1}, it turns out that the function $f_{N_o}(x,y)$ is only a rough estimator of the hypothetical PDF. This property is a consequence of the fact that experimental information $I$ and redundancy $R$ have equal weights in the cost function $C=R-I$.
\begin{figure}
\centering
\includegraphics[width=3.375in]{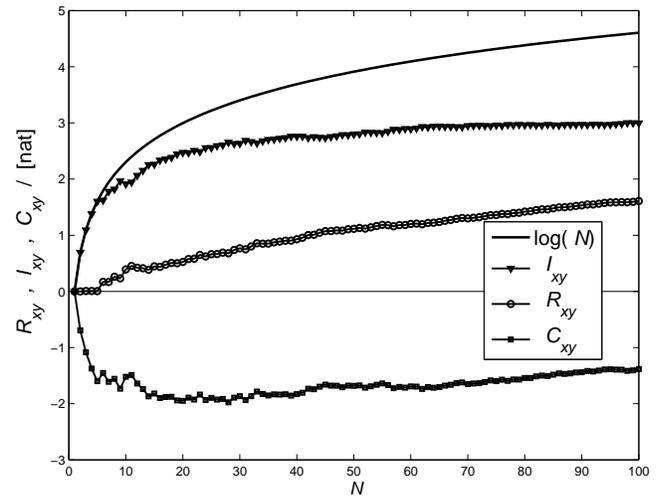}
\caption{Dependence of information statistics on the number $N$ of data points as determined based on integration.}
\label{IRCN}
\end{figure}

Figs.\,\ref{IRCaN} and \ref{IRCN} indicate that experimental information $I$ converges with increasing $N$ to a certain limit value from which the complexity of the phenomenon under investigation can be approximately estimated as $K\approx \exp I_{N_{max}}$. In our case we get the estimate $K\approx 21$. The number of non--overlapping scattering distributions that represent the PDF of the observed phenomenon is thus slightly smaller than the proper number $N_o$ of experiments needed for its exploration.

\section{Conclusions}

From the statistics introduced in the previous articles \cite{ig1,ig2} based on information entropy, we have here derived an approximate formula for the calculation of redundancy $R$ of experimental data. It is important that this formula does not include the integral by which the information entropy is defined. This makes feasible a simplified and fairly good estimation of redundancy and, with it, the related experimental information and cost function. The advantage of the approximate calculation becomes outstanding in multivariate cases because multiple integration is not needed there. A serious obstacle for the application of the concept of experimental information and redundancy of data can thus be avoided. Efficient estimation of the experimental information and cost function, and with them the determined complexity of the phenomenon and the proper number of experiments needed for its exploration, could be considered valuable in planning experimental work. In addition, the complexity $K$ or the proper number $N_o$ could be applied in the field of neural networks \cite{gs,ha} to determine the appropriate number of cells needed to deal with a certain phenomenon. 

\noindent{\bf Acknowledgment}
\newline\noindent This work was supported by the Ministry of Higher Education, Science and Technology of the Republic of Slovenia and EU -- COST.

\end{document}